\documentclass[twocolumn,showpacs,aps,prl,superscriptaddress]{my_revtex4}

\usepackage{graphicx}
\usepackage{dcolumn}
\usepackage{amsmath}
\usepackage{epsfig}

\input babarsym
\newcommand{\bi}{\begin{itemize}}
\newcommand{\ei}{\end{itemize}}
\newcommand{\ben}{\begin{enumerate}}
\newcommand{\een}{\end{enumerate}}
\newcommand{\bc}{\begin{center}}
\newcommand{\ec}{\end{center}}
\newcommand{\bt}{\begin{table}}
\newcommand{\et}{\end{table}}
\newcommand{\be}{\begin{equation}}
\newcommand{\eeq}{\end{equation}}
\newcommand{\ba}{\begin{eqnarray}}
\newcommand{\ea}{\end{eqnarray}}

\newcommand{\vs}{\vspace}
\newcommand{\la}{\ifmmode {\leftarrow} \else {$\leftarrow$}\fi}
\newcommand{\Ra}{\ifmmode {\Rightarrow} \else {$\Rightarrow$}\fi}
\newcommand{\La}{\ifmmode {\Leftarrow} \else {$\Leftarrow$}\fi}
\newcommand{\Lra}{\ifmmode {\Longrightarrow} \else {$\Longrightarrow$}\fi}
\newcommand{\Lla}{\ifmmode {\Longleftarrow} \else {$\Longleftarrow$\fi}}
\newcommand{\Llra}{\ifmmode {\Longleftrightarrow} \else {$\Longleftrightarrow$\fi}}
\newcommand{\Lk}{\ifmmode {{\cal L}} \else {${\cal L}$}\fi}
\newcommand{\Wt}{\ifmmode {{\cal W}} \else {${\cal W}$}\fi}
\newcommand{\Br}{\ifmmode {{\cal B}} \else {${\cal B}$}\fi}
\newcommand{\N}{\ifmmode {{\cal N}} \else {${\cal N}$}\fi}
\newcommand{\G}{\ifmmode {{\cal G}} \else {${\cal G}$}\fi}
\newcommand{\E}{\ifmmode {{\cal E}} \else {${\cal E}$}\fi}

\newcommand{\tBz}{\ifmmode {\tau_{\Bz}} \else {$\tau_{\Bz}$ }\fi }
\newcommand{\tBp}{\ifmmode {\tau_{\Bu}} \else {$\tau_{\Bu}$ }\fi }
\def\nubar    {\kern 0.18em\overline{\kern -0.18em \nu}{}\xspace}
\def\nulb     {\ensuremath{\nubar_\ell}\xspace}
\newcommand{\BtoDs}{\mbox{$\Bzb\rightarrow D^{*+} \ell^- \nulb$}}

\newcommand{\BtoDss}{\mbox{$B\rightarrow D^{*+} \pi \ell^- \nulb$}}

\newcommand{\psoft}{\ifmmode {\pi_s^+} \else {$\pi_s^+$}\fi }
\newcommand{\dm}{\ifmmode {\Delta M} \else {$\Delta M$}\fi}
\newcommand{\plab}{\ifmmode{p} \else {$p$}\fi}
\newcommand{\ks}{\ifmmode{k^*} \else {$k^*$}\fi}
\newcommand{\om}{\ifmmode{w} \else {$w$}\fi}
\newcommand{\omt} {\ifmmode {\tilde{w}} \else {$\tilde{w}$} \fi}
\newcommand{\mnusq}{\ifmmode{{M_\nu}^2} \else {${M_{\nu}}^2$}\fi} 
\newcommand{\DTau}{\ifmmode {\Delta \tau} \else {$\Delta \tau$}\fi}
\newcommand{\ggcc}{\ifmmode {GeV^2/c^4} \else {$GeV^2/c^4$}\fi}
\newcommand{\TBY}{\ifmmode{\theta_{\Bz, D^*\ell}} \else {$\theta_{\Bz, D^*\ell}$} \fi}
\newcommand{\Aone}{\ifmmode {{\cal A}_1} \else {${\cal A}_1$}\fi}
\newcommand{\rha}{\ifmmode{\mbox{\rho^2_{{\cal A}_1}}} \else {\mbox{$\rho^2_{{\cal A}_1}$}}\fi}
\def\BpBm {\ensuremath{B^+ {\kern -0.16em \Bub}}}

\newcommand{\BABARPubYear}    {04}
\newcommand{\BABARPubNumber}  {018}

\newcommand{\SLACPubNumber} {10591}
\newcommand{\LANLNumber} {0408027}

\def\figurebox#1#2#3{%
    \def\arg{#3}%
    \ifx\arg\empty
    {\hfill\vbox{\hsize#2\hrule\hbox to #2{\vrule\hfill\vbox to #1{\hsize#2\vfill}\vrule}\hrule}\hfill}%
    \else
    {\hfill\epsfbox{#3}\hfill}%
    \fi}

\begin{document}

\preprint{\babar-PUB-\BABARPubYear/\BABARPubNumber} 
\preprint{SLAC-PUB-\SLACPubNumber} 

\begin{flushleft}
\babar-PUB-\BABARPubYear/\BABARPubNumber\\
SLAC-PUB-\SLACPubNumber\\
hep-ex/\LANLNumber\\[10mm]
\end{flushleft}

\vspace*{-0.1cm}
\title{
{\large \bf Measurement of the \mbox{{\boldmath $\Bzb \ra D^{*+} \ell^- \bar{\nu}_{\ell}$}} Decay Rate and {\boldmath \Vcb}}}

%
\author{B.~Aubert}
\author{R.~Barate}
\author{D.~Boutigny}
\author{F.~Couderc}
\author{J.-M.~Gaillard}
\author{A.~Hicheur}
\author{Y.~Karyotakis}
\author{J.~P.~Lees}
\author{V.~Tisserand}
\author{A.~Zghiche}
\affiliation{Laboratoire de Physique des Particules, F-74941 Annecy-le-Vieux, France }
% ...
\collaboration{The \babar\ Collaboration}
\noaffiliation

\date{\today}
\memory{Paolo Poropat}%

\begin{abstract}
We present a measurement of the Cabibbo-Kobayashi-Maskawa matrix element \Vcb\ 
based on a sample of  about 53,700 
\mbox{$\Bzb \rightarrow D^{*+} \ell^- \bar{\nu}_{\ell}$} decays 
observed by the \babar\  detector.  We obtain the branching fraction averaged 
over $\ell = e,\mu$,
${\cal B}(\Bzb \rightarrow D^{*+} \ell^- \bar{\nu}_{\ell}) = (4.90 \pm 0.07\mathrm{(stat.)}^{+0.36}_{-0.35}\mathrm{(syst.)})\%$.
We measure the differential
decay rate as a function of $w$, the relativistic boost $\gamma$ of the
$D^{*+}$ in the ${\Bzb}$ rest frame.
By extrapolating $d\Gamma/dw$ to the kinematic limit $w \rightarrow 1$, 
we extract the product of \Vcb and the axial form factor ${\cal A}_1(w=1)$.
We combine this measurement with a lattice QCD calculation of ${\cal A}_1(w=1)$ to determine
$
\nonumber |V_{cb}| = (38.7 \pm 0.3 \mathrm{(stat.)} \pm 1.7 \mathrm{(syst.)} ^{+1.5} _{-1.3} \mathrm{(theory)})\times 10^{-3} . 
$
\end{abstract}


\maketitle
In the Standard Model of electroweak interactions,
the Cabibbo-Kobayashi-Maskawa (CKM) matrix describes
the flavor mixing among quarks and determines the strength of \CP\-violation by a single non-trivial weak phase. 
The CKM matrix element $V_{cb}$ measures the weak coupling of the $b$ to the $c$ quark. 
In this Letter, we present  measurements of the branching fraction ${\cal B}(\BtoDs)$~\cite{footnote1}  and \Vcb. 
The rate for this weak decay is proportional to $\Vcb^2$ and is influenced by strong interactions 
through form factors, which are not known a priori.
In the limit of infinite $b$-quark and $c$-quark masses, these form factors are determined by a single
Isgur-Wise function~\cite{Isgurwise}. The value of this function when the \dsp\ is at rest relative to
the \Bzb\ has been computed for finite $c$- and $b$-quark masses using lattice QCD \cite{Hashimoto}.

In this analysis, we measure the differential decay rate $d\Gamma/d\om$, 
where \om\ is the product of the four-velocities of the \Bzb\ and \dsp , and corresponds
to the relativistic boost $\gamma$ of the \dsp\ in the \Bzb\ rest frame.
We extrapolate the rate to the zero-recoil limit \om=1, and use the theoretical result for the form factor 
there \cite{Hashimoto} to extract \Vcb. 

The analysis is based on a data sample of 79~\invfb\ recorded  on the \FourS\  resonance and 9.6~\invfb\ recorded 40~MeV
below it, with the \babar\ detector \cite{babar} at the PEP-II asymmetric-energy \epem\ collider. 
We use samples of GEANT Monte Carlo (MC) simulated events that correspond to about three times the data sample size. 

The momenta of charged particles are measured by a tracking system consisting of a five-layer silicon vertex tracker (SVT) 
and a 40-layer drift chamber (DCH), operating in a 1.5-T solenoidal magnetic field.
Charged particles of different masses  are distinguished by their energy loss in the 
tracking devices and by a ring-imaging Cherenkov detector. Electromagnetic showers from electrons and 
photons are measured in
 a CsI(Tl) calorimeter. Muons are identified in a set of resistive plate chambers inserted in the iron flux-return yoke of the 
magnet.

We select events that contain a $\dsp$ candidate
and an oppositely charged electron or muon with momentum $1.2<p_{\ell}<2.4~\gevc$.
(Unless explicitly stated otherwise, momenta are measured in the \FourS\ rest frame, which
does not coincide with the laboratory frame, due to the boost of the PEP-II beams.)
In this momentum range, the electron (muon) efficiency is about 90\% (60\%) and the hadron misidentification 
rate is typically 0.2\% (2.0\%). We select \dsp\ candidates in the momentum range
$0.5 < p_{D^*} < 2.5~\gevc$ in the channel $\dsp \ra \Dz \psoft$, with the \Dz\ decaying to $K^-\pi^+,~K^-\pi^+\pi^-\pi^+$, or $K^-\pi^+\pi^0$.
The charged hadrons of the \Dz\ candidate are fit to a common vertex and the candidate is
rejected if the fit probability is less than $0.1\%$. We require the invariant mass of the hadrons to be within
17~\mevcc\ of the \Dz\ mass for the decays to only charged particles, and 34~\mevcc\ for $K^-\pi^+\pi^0$ decays. 
For $\Dz \ra K^-\pi^+\pi^0$, we accept only candidates from portions of the Dalitz plot where the square of the decay 
amplitude, as determined by Ref.~\cite{Dalitz}, is at least 10$\%$ of the maximum it attains anywhere in the plot.
For the pion from \dsp\ decay, \psoft, the momentum in the laboratory frame must be less than 450~\mevc , 
and the transverse momentum greater than 50~\mevc .
Finally, the lepton, $\psoft$, and \Dz are fit to a common vertex with a beam-spot constraint, 
and the probability for this fit is required to exceed 1\%.

In semileptonic decays, the presence of an undetected neutrino complicates the separation of the
signal from background. We compute a kinematic variable with considerable power to reject background
by determining,
for each $B$-decay candidate, 
the cosine of the angle between the momentum of the \Bzb\ and of the $\dsp\ell^-$ pair,
under the assumption that only a massless neutrino is missing:
\begin{equation*}
 \cos\TBY = \frac{2E_{\Bzb} E_{D^*\ell} - M^2_{\Bzb} - M^2_{D^*\ell} } { 2 p_{\Bzb} p_{\raisebox{-0.3ex}{\scriptsize $D^*\ell$}} }.
\end{equation*}
This quantity constrains the direction of the  \Bzb\  to lie along a cone whose axis is the
direction of the $\dsp \ellm$ pair, but with an undetermined 
azimuthal angle about the cone's axis. The value of $w$ varies with this azimuthal angle; 
we take the average of the minimum and maximum values as our estimator \omt for $w$. 
This results in a resolution of $0.04$ on \om .
We divide the sample into 10 bins in \omt\ from 1.0 to 1.5, with the last bin extending to the 
kinematic limit of 1.504.

The selected events are divided into six subsamples, 
corresponding to the two leptons and the three \Dz\ decay modes.
In addition to signal events, each subsample contains  backgrounds from  six different sources: combinatoric 
(events from $\BB$ and continuum in which at least one of the hadrons assigned to the \dsp\ does not
originate from \dsp\ decay); continuum ($\dsp \ell^-$ combinations from $\epem \ra c\bar{c}$); 
fake leptons (combined with a true \dsp); uncorrelated background 
($\ell$ and \dsp\ produced in the decay of two different $B$ mesons); events from \BtoDss\ decays ; and
correlated background events due to the processes $\Bzb \ra \dsp \bar{\nu} \tau^-,~\tau^- \ra \ellm X$ 
and $\Bzb \ra \dsp X_c, ~X_c \ra \ellm Y$. We estimate correlated background (which amounts to 
less than 0.5\% of the selected candidates) from Monte Carlo simulation
based on measured branching fractions \cite{PDG}, while we determine all the others from the data.
Except for the combinatoric background, 
all other background sources exhibit a peak in the $\dm = M_{\dsp} - M_{\Dz}$ distribution, where $M_{\dsp}$ and $M_{\Dz}$ are the measured \dsp\ and \Dz\ candidate masses. 

We determine the composition of the subsamples in each \omt\ bin in two steps.
First we estimate the amount of combinatoric, continuum, and fake-lepton background
by fitting the \dm\ distributions in the range $0.139<\dm<0.165~\gevcc$ simultaneously 
to three sets of events: data recorded on resonance, data taken below the \FourS (thus containing only continuum background),
 and data in which tracks that fail very loose lepton-selection criteria are taken as surrogates for fake leptons.
The distributions are fit with the sum of two Gaussian functions with a common mean and different widths 
to describe $\dsp \ra \Dz \psoft$ decays and empirical functions, based on the simulation, 
for the combinatoric background. The four parameters of the Gaussian functions 
are common, while 
the fraction of peaking events and the parameters describing the combinatoric background differ for the signal, 
off-peak, and fake-lepton samples.

\begin{figure}[!t]
\begin{center}
\begin{tabular}{l}
\vs{-0.5cm}\includegraphics[height=5.2cm]{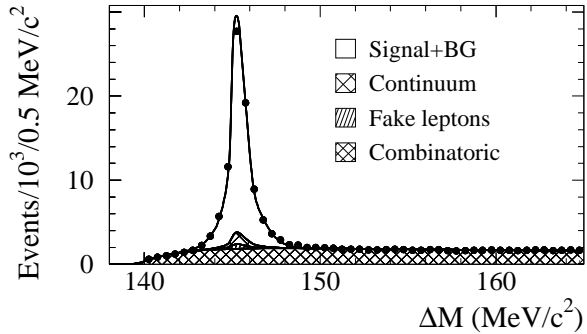}
\end{tabular}
\vs{-0.3cm}
\caption{Yields of on-resonance data (points) and the results of the fit (line)
to the \dm\  distribution, with contributions from continuum, fake-lepton, and 
combinatoric-\dsp\ backgrounds summed over all \omt\ bins. }
\vs{-0.5cm}
\label{f:dm}
\end{center}
\end{figure}

Since the \dm\ resolution depends on whether or not the \psoft~ track 
is reconstructed only in the SVT or in the SVT and DCH, the fits are performed separately 
for these two classes of events.
We rescale the number of continuum and fake-lepton events in the  mass range 
$0.143 < \dm < 0.148~\gevcc$, based on the
relative on- and off-resonance luminosity and measured hadron misidentification probabilities. 
In the subsequent analysis we
fix the fraction of combinatoric, fake-lepton, and continuum events in each \omt\ bin to the values so obtained.
Figure 1 shows the \dm\ fit results for the on-resonance data.

In a second step, we fit the $\cos\TBY$ distributions in the range 
$-10 < \cos\TBY < 5$ and determine the signal contribution and the normalization of the 
uncorrelated and \BtoDss\ backgrounds. 
Neglecting resolution effects, signal events meet the obvious constraint $|\cos\TBY|<1$, while \BtoDss\ 
events extend below $-1$, and uncorrelated background events are spread over 
the entire range considered.

\begin{figure}[!t]
\begin{center}
\begin{tabular}{l}
\vs{-0.5cm}\includegraphics[height=5.2cm,width=8.2cm]{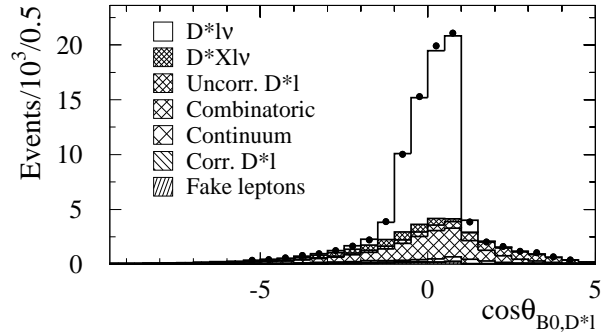}
\end{tabular}
\vs{-0.3cm}
\caption{ Yields of on-resonance data (points) and the results of the fit (histograms) to the $\cos\TBY$  
distribution, summed over all \omt\ bins.}
\vs{-0.5cm}
\label{f:ctBY}
\end{center}
\end{figure}

We perform the fit separately for each \omt\ bin, with the individual shapes for the signal and for each of the 
six background sources taken  from MC simulation, specific for each of the six subsamples. 
Signal events are generated with the form-factor parameterization of Ref.~\cite{Caprini}, tuned to
the results from CLEO~\cite{CLEOff}. Radiative decays (\BtoDs $\gamma$) are modeled by PHOTOS \cite{PHOTOS}
and treated as signal. $\B \to D^{**} \ell \nu$ decays involving orbitally 
excited charm mesons are generated according to the ISGW2 model~\cite{IGSW}, and decays with nonresonant charm states  
are generated following the prescription in Ref.~\cite{Goity}. 
To reduce the sensitivity to statistical fluctuations we require that the ratio of 
 \BtoDss\ and of uncorrelated background to the signal be the same for all three \Dz\ decay modes and for the electron and muon samples.
Fit results are shown in Fig.~\ref{f:ctBY}. 
In total, there are 70,822 events in the range $|\cos\TBY|<1.2$. 
The average fraction of these events that are signal is $(75.9\pm0.3)$\%, where the error is only statistical.

To extract \Vcb, we compare the signal yields to the expected differential decay rate 
\begin{equation*}
\frac{{\rm d}\Gamma}{{\rm d}\om}
 = \frac{G^2_F}{48\pi^3} M_{\dsp}^3 (M_{\Bzb}-M_{\dsp})^2 {\cal G}(w){\cal F}(w)^2 \Vcb^2 ,
\end{equation*}
where
\begin{equation*}
{\cal G}(\om)= \sqrt{\om^2-1}(\om+1)^2 \left( 1 + 4 \frac{\om}{\om+1} \frac{1-2\om r+r^2}{(1-r)^2}\right)
\end{equation*}
is a phase-space factor, $r= M_{\dsp}/M_{\Bzb}$. We parameterize the form factor ${\cal F}(w)$ with a Taylor
expansion:
\begin{equation*}
{\cal F}(w) \approx {\cal F}(1) ( 1 -\rho_{\cal{F}}^2 (\om-1) + c (\om -1)^2 ),
\end{equation*}
where we neglect terms of order greater than two in $(\om-1)$. 
We fit the data to determine ${\cal F}(1) \Vcb$, $\rho_{\cal{F}}$
and $c$.

Dispersion relations inspired by QCD can be used to constrain the shape of the form factor 
and reduce the number of parameters to be determined~\cite{Lebed,Caprini}.
Therefore we consider also the parameterization proposed in Ref.~\cite{Caprini}, 
which relates ${\cal F}(w)$ to the axial-vector
form factor ${\cal A}_1(\om)$ according to the following expression:
\begin{eqnarray*}
{\cal F}(w)^2{\cal G}(w) = \Aone(\om)^2\sqrt{\om-1} (\om+1)^2 \left\{ 2 \left[\dfrac{1-2\om r+r^2}{(1-r)^2}\right] \right. \\
\left. \times \left(1+R_1(\om)^2 \dfrac{\om-1}{\om+1}\right) + \left[1+(1-R_2(\om)) \dfrac{\om-1}{1-r}\right]^2 \right\},
\end{eqnarray*}
where $R_1(\om) \approx R_1(1) -0.12(\om-1) +0.05(\om-1)^2$, 
$R_2(\om) \approx R_2(1) +0.11(\om-1) -0.06(\om-1)^2$, and 
we use the values $R_1(1)=1.18 \pm 0.32$ and $R_2(1)=0.71 \pm 0.21$ 
measured by CLEO \cite{CLEOff}. Using dispersion relations we express the ratio \Aone(\om)/\Aone(1) as a function
of a single unknown parameter \rha:
\begin{eqnarray*}
\frac{{\cal A}_1(\om)}{{\cal A}_1(1)} \approx  1-8\rho^2_{{\cal A}_1} z +(53\rho^2_{{\cal A}_1}-15)z^2-(231\rho^2_{{\cal A}_1}-91)z^3 ,
\end{eqnarray*}
where $z = (\sqrt{\om+1}-\sqrt{2})/(\sqrt{\om+1}+\sqrt{2})$.
It must be noted that, for $\om \ra 1$, $\Aone(\om) \ra {\cal F}(\om)$, so we expect $\Aone(1) \approx {\cal F}(1)$. 

We perform a least-squares fit of the sum of the observed signal plus background yields to the expected yield in the ten 
bins in \omt . We define for each of the six data subsamples 
\begin{equation*}
\chi^2 = \sum_{i=1}^{10} \frac{(N^i_{\rm data} - N^i_{\rm bk}- \sum_{j=1}^{N^i_{\rm MC}} W^{i}_j )^2}
                                { N^i_{\rm data} + {\sigma^i_{\rm bk}}^2 +  \sum_{j=1}^{N^i_{\rm MC}} {W^{i}_j}^2},
\end{equation*}
where $N^i_{\rm data} $ is the number of observed events in 
the $i^{th}$ bin; $N^i_{\rm bk}$ and $\sigma^i_{\rm bk}$ are the number of estimated background events and its error. 
The backgrounds are fixed to the estimated rates.  
The expected signal yield is calculated at each step of the minimization from the reweighted sum of $N^i_{MC}$ simulated events.
  Each weight is the product of four weights,
$W_j^i = W^{\cal L} \, W^{\epsilon,i}_j \, W^{\cal S} \, W^{ff,i}_j$. 
The factors $W^{\cal L}$, $W^{\epsilon,i}_j$ do not vary during the minimization, while
the terms $ W^{\cal S}$, $W^{ff,i}_j$ depend on parameters which are determined by the fit, and vary at 
each step of the minimization. 

The first factor $W^{\cal L}$ accounts for relative normalization of the data and MC samples, 
and is common to all subsamples.  $W^{\cal L}$ depends on the total number of
\BB\ events, $N_{\BB} = (85.9 \pm 0.9)\times 10^6 $, 
on the fraction of \BzBzb events, $f_{00} =0.489\pm 0.012$~\cite{PDG},
on the branching fraction ${\cal B}(\dsp \ra \Dz \pi^+)= 0.677\pm 0.005$~\cite{PDG},
and on the \Bz\ lifetime $\tau_{\Bz} = 1.536 \pm 0.014 $~ps~\cite{PDG}. 
$ W^{\epsilon,i}_j$ accounts for differences in reconstruction and 
particle-identification efficiencies predicted by the Monte Carlo simulation 
and measured with data, as a function of particle momentum.
Only the \psoft\ tracking efficiency varies significantly with $\omt$.

The weight $W^{\cal S}$ 
accounts for  potential small differences in 
efficiencies for the six data subsamples and allows for adjustments 
of the \Dz\ branching fractions, properly dealing with the correlated systematic
uncertainties.
It is the product of several scale factors that are floating parameters in the fit, each constrained to an expected value with a corresponding experimental error. For instance, to account for the uncertainty in the 
multiplicity-dependent tracking efficiency, we introduce a factor $W^{\cal S}_{trk}=1+N_{trk}\delta_{trk}$, 
where $N_{trk}$ is the number of charged tracks in the $\dsp \ell^-$ candidates in each sample
 and $\delta_{trk}$ is constrained to zero within the 
estimated uncertainty in the single-track efficiency: $\pm0.8\%$.
Similarly, correction factors are introduced to adjust lepton, kaon, and 
$\pi^0$ efficiencies, and \Dz\ branching fractions, taking into account correlations. 

The fourth factor, $W^{ff,i}_j$, adjusts the fitted decay distribution relative to the 
one used in the generation of the MC events. This term depends on \Vcb\ and on the shape parameters. 
It is a function of \om\ and is determined for each simulated event at each step of the fit.

\begin{figure}[!t]
\begin{center}
\begin{tabular}{l}
\vs{-1.88cm}\includegraphics[height=5.0cm]{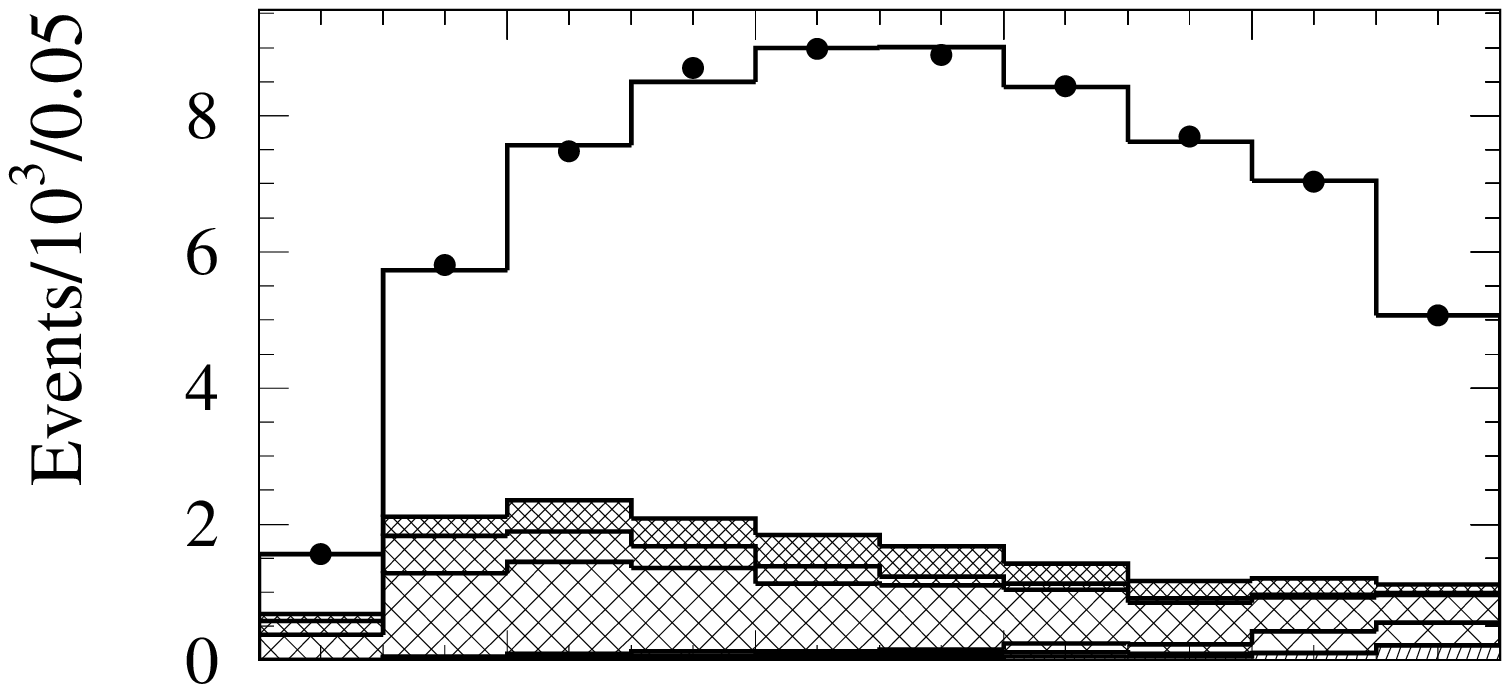}\\
\vs{ -0.3cm}\includegraphics[height=5.0cm]{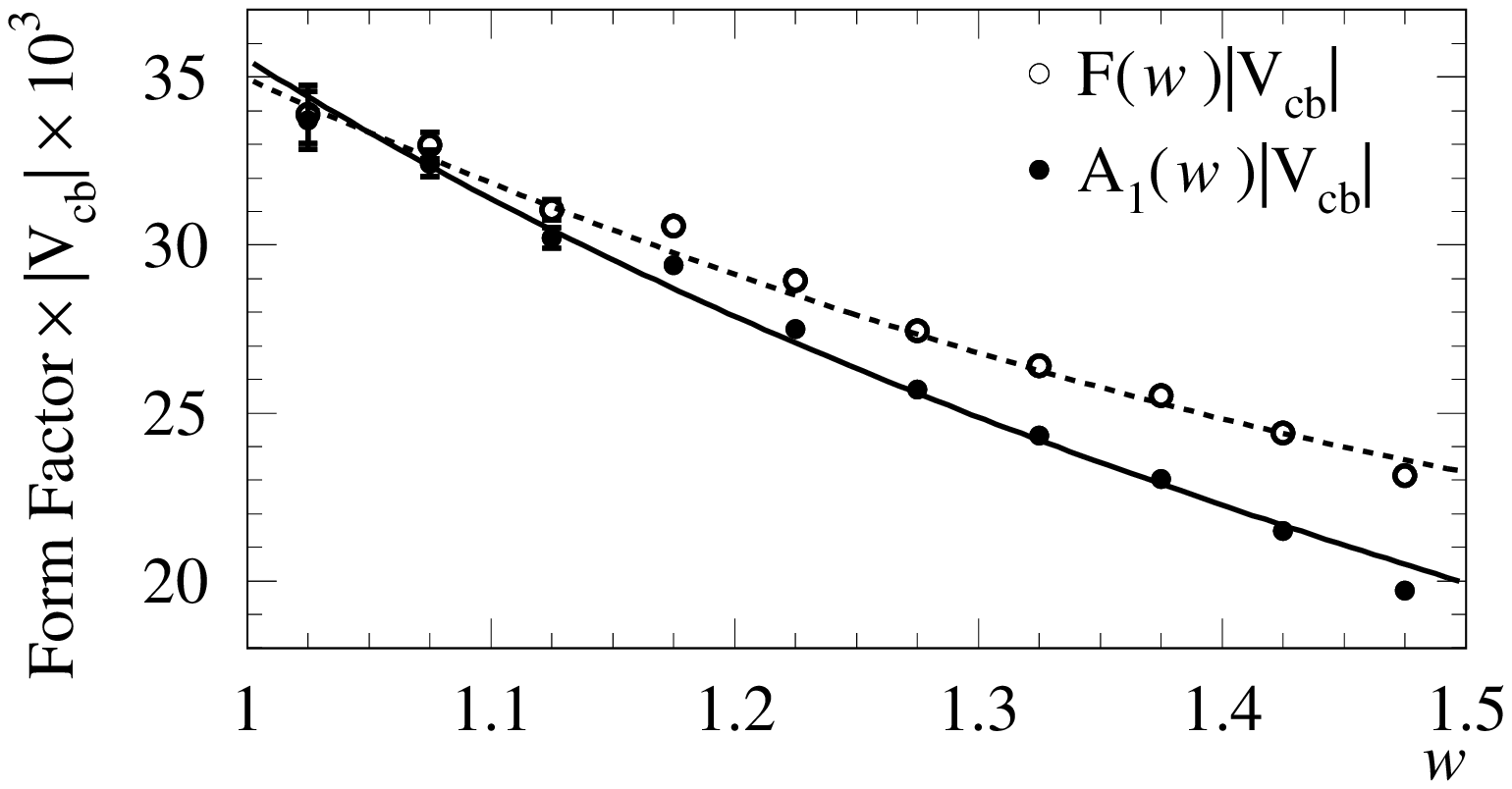}
\end{tabular}
\vs{-0.7cm}
\caption{Results of the fit as a function of \omt\ compared to data.
Top:  the observed \omt distribution (points) compared to the fit result;
signal and background contributions are indicated using the same shading as in Fig.\ref{f:ctBY}.
Bottom: the form factor parameterizations with fitted parameters compared  
to the background- and efficiency-corrected data. The solid (dotted) line corresponds to the $\Aone(\om)$ 
(${\cal F}(\om)$) parametrization, and is to be compared to the filled (open) data points.}
\vs{-0.5cm}
\label{f:fit}
\end{center}
\end{figure}

Figure~\ref{f:fit} (top) compares the observed signal and background yields, summed over all six
subsamples, with the result of the fit. Figure~\ref{f:fit} (bottom) illustrates the extrapolation to $w=1$ for 
the two form-factor parameterizations.  
The numerical values obtained for the two different form-factor parameterizations are listed in Table \ref{t:results}. 
For both fits, the $\chi^2$ per degree of freedom is satisfactory, and the scale factors introduced to allow adjustments of 
the efficiencies and branching fractions deviate from their default values by less than one 
standard deviation.

\begin{table}[htb]
\caption{Results of the fits to $d\Gamma / d \omt$ for the two parameterizations of the form factor.
The errors stated include statistical error of the data and MC as well as uncertainties due to tracking, 
particle identification, and \Dz\ branching fractions that are directly assessed in the fit procedure.}
\begin{center}
\begin{tabular}{lcccc} \hline\hline
                      & ${\cal A}_1(1) \Vcb\times 10^3$ & $\rho^2$ & $c$ &  $\chi^2/\rm ndf$ \\  \hline
${\cal F}$ & $35.0\pm 0.9$ & $0.95\pm 0.09$    & $0.54\pm0.17$           &  $67/57$  \\
${\cal A}_1$ & $35.5\pm 0.8$ & $1.29\pm 0.03$    & -                 &  $69/58$  \\
\hline\hline
\end{tabular}
\end{center}
\label{t:results} 
\end{table}

In Table II we present a summary of the statistical and systematic 
uncertainties. From the fit to the \omt distribution we obtain errors that 
combine the statistical error with systematic errors introduced by 
the uncertainties in scale factors. We separate the various 
contributions in the following way: first, we extract the statistical 
errors by fixing all scale factors to their fitted values. 
The systematic errors due to the uncertainties in a given scale 
factor is extracted from a separate fit in which this scale 
factor is fixed. We take the square root of reduction in the square of the fit errors 
as a measure of the contribution of the particular scale factor to 
the overall error in the fit parameters. 

We then assess the individual contributions to the systematic error due to other input quantities by varying 
their values by their estimated uncertainties and adding in quadrature the resulting changes to the fit parameters.
The uncertainties in the lifetime $\tau_{\Bz} $, the \FourS\ and \dsp\ branching fractions, and overall normalization 
are independent of $w$ 
and thus do not affect the shape of the form factor. The uncertainty introduced by the  vertex reconstruction is 
common to all samples and independent of $w$. It is determined by comparing the event samples with and without cuts
on the vertex probability.
The error induced by the cut on the decay amplitude for the $K^-\pi^+\pi^0$ decay is determined by varying that cut. 

A major source of uncertainty is the reconstruction efficiency for the low-momentum pion from 
the \dsp\ decay, since it is 
highly correlated with the \dsp\ momentum and thereby with $w$.
We determine the tracking efficiency for high-momentum tracks comparing the independent 
information from SVT and DCH. We compute the efficiency for low-momentum tracks reconstructed in the SVT alone from the angular distribution of the \psoft\ in the \dsp\ rest frame. We use a large set of $\dsp\ra\Dz\psoft$, $\Dz\ra K^-\pi^+$ decays selected from 
generic hadronic events. For fixed values of the \dsp\ momentum, we compare the observed angular distribution
to the one expected for the decay of a vector meson to two pseudoscalar mesons. 
We define the relative efficiency as the ratio of the observed to the expected distribution and 
parameterize its dependence on the laboratory momentum of the \psoft. The study is performed in several bins
of the polar angle of the detector. 
We perform the measurement in the data and in the
simulation, and we find that the functions parameterizing the efficiency are consistent within the statistical 
errors. To assess the systematic uncertainty on \Vcb , we vary the parameters of the 
efficiency function by their uncertainty, including correlations. We  
add in quadrature the uncertainty in the 
absolute scale, as determined using high-momentum tracks reconstructed in both the SVT and the DCH. 
We obtain a systematic error of $\pm1.1 \%$ on \Vcb.



\begin{table}[hbt]
\caption{Summary of uncertainties.}
\begin{center}
\begin{tabular}{lccc}
\hline\hline
Source of Uncertainty & $\delta(\Aone(1)\Vcb)$($\%$)   & $\delta$\rha  & $\delta {\cal B}$($\%$) \\
\hline

Data and MC statistics                         & 0.7   & 0.03  & 1.4 \\

${\cal B}(D^{0}\ra\ K^-\pi^+)$            & 1.1   & -     & 2.2 \\
${\cal B}(D^{0}\ra\ K^-\pi^+\pi^-\pi^+ )$ & 0.4   & -     & 0.8 \\
${\cal B}(D^{0}\ra\ K^-\pi^+\pi^0 )$      & 0.5   & -     & 1.0 \\
Particles identification                  & 1.1   & -     & 2.2 \\
Tracking \& $\pi^0$ reconstr.             & 1.3   & -     & 2.6 \\
\hline
Partial Sum                     	  & 2.2   & 0.03  & 4.5 \\
\hline\hline

\Bz\ lifetime                           & 0.5   & -     & -   \\
Number of $B\overline B$                & 0.6   & -     & 1.2 \\
${\cal B}(D^{*+}\ra\Dz\pip)$            & 0.4   & -     & 0.7 \\
${\cal B}(\FourS\ra\BzBzb)$             & 1.2   & -     & 2.5 \\
$D^{*+}\ell^-$ vertex efficiency    & 0.5   & -     & 1.0 \\
$\pi_s$ efficiency                      & 1.1   & 0.01  & 1.9 \\
$D^{*}\pi\ell\nu$ sample composition              & 1.8   & 0.06  & 2.0 \\
\B\ momentum                    & 0.3   & -     & 0.7 \\
Radiative corrections           & 0.2   & 0.01  & 0.4 \\
$\cos\TBY$ $\&$ \omt\ fit method& 0.8   & 0.02  & 1.6 \\
$R_1(1)$ and $R_2(1)$                 & ${\phantom 1}^{+2.9}_{-2.6}$   & 0.26  &  ${\phantom 1}^{+3.9}_{-3.3}$ \\
\hline\hline
Total Error                     & {\boldmath ${{\phantom 1}^{+4.6}_{-4.4}}$}& {\bf 0.27}& 
                                  {\boldmath ${{\phantom 1}^{+7.4}_{-7.1}}$} \\
\hline\hline
\end{tabular}
\end{center}
\label{t:syst}
\end{table}

The largest error in the background subtraction is due to the uncertainty in the composition and form factors of the 
 $\dsp \pi \ell^- \bar{\nu}_{\ell}$ decays.
We consider twelve different $\dsp \pi$ states, narrow and wide, as well as nonresonant $\dsp \pi$. 
To assess the impact of these decays on the fit we repeat the analysis assuming that only one mode at a time 
populates the
whole sample, and then 
take as the systematic error half the difference between the maximum and minimum fitted parameters.

We assess the effect of the uncertainty in the average \Bzb\ momentum, as determined from 
a sample of fully reconstructed hadronic $B$ decays on the fit results. 
We take into account an uncertainty of $\pm 30\%$ in the emission rate of the radiative photons 
predicted by PHOTOS~\cite{PHOTOS}. 

We also assess the impact of changes in the bin size on the fits to the $\cos\TBY$ and \omt\ distributions. 

There are several uncertainties related to the form factors and their parameterization. 
The form factor ratios $R_1$ and $R_2$ affect the lepton momentum spectrum and thus the differential 
decay rate as a function of $w$, as well as the fraction  of events satisfying the lepton momentum requirements. 
We assess these effects by varying $R_1$ and $R_2$ within the measurement errors~\cite{CLEOff}, 
taking into account their correlation.  
As a consistency check, we compare the measured momentum spectra of the \dsp\ and leptons with the 
spectra expected from the fit results. 
We find very good agreement for the \dsp, but the lepton spectrum favors a larger value for $R_1$, 
though one consistent with the available measurement.

If we fit separately $e$ and $\mu$ samples, we find exactly the same value for \rha.
The values of $\Aone \Vcb$ ,  $(35.8 \pm 0.5) \times 10^{-3 }$ and $(35.0 \pm 0.5) \times 10^{-3 }$ respectively,
differ by 1.2 standard deviation. 
  
The value of $c$, given in Table~\ref{t:results}, shows that
the data disfavor a purely linear dependence of ${\cal F}$ on $\om$, by almost three standard deviations.
The fits for the two different parameterizations of the \om\ dependence of the form factors are consistent at $\om=1$. 
We choose 
$\Aone(1) \Vcb\ = (35.5\pm 0.3 \pm 1.6 ) \times 10^{-3}$, and
\rha = $1.29 \pm 0.03 \pm 0.27$,
where the errors listed refer to the statistical, and the systematic uncertainties.
The correlation between $\Aone(1) \Vcb$ and \rha\ is 0.56, taking into account statistical and systematic errors.
A recent lattice calculation~\cite{Hashimoto} (including a QED correction of 
0.7\%) gives $\Aone(1) = {\cal F}(1) = 0.919 ^{+0.030}_{-0.035}$, with which we 
obtain
\begin{eqnarray} 
\nonumber  |V_{cb}| = (38.7\pm 0.3 \pm 1.7  {\phantom 1 } ^{+1.5} _{-1.3} ) \times 10^{-3},
\end{eqnarray}
where the first error is statistical, the second is systematic, 
and the third  reflects the uncertainty in $\Aone(1)$.
Integrating over the fitted \omt\ distribution these parameters result in the branching fraction 
${\cal B} (\Bzb \ra \dsp \ell^- \bar{\nu}_{\ell}) = (4.90 \pm 0.07 ^{+0.36}_{-0.35})\%$, 
where the errors are the statistical and systematic uncertainties.

In summary, we have measured the CKM parameter \Vcb\ and the exclusive branching fraction for 
$\Bzb \ra \dsp \ell^- \bar{\nu}_{\ell}$ with high precision.  
The result for \Vcb\ is consistent with another \babar\ measurement
 based on lepton and hadron spectra from inclusive semileptonic $B$-meson decays~\cite{bbrVcb}, 
$\Vcb = (41.4\pm0.4(\mathrm{stat.})\pm0.4(\mathrm{exp.})\pm 0.6(\mathrm{theory}))\times 10^{-3}$.
The results for \Vcb\ and the branching fraction are also consistent with earlier 
measurements~\cite{vcb} based on the technique employed here, except for those 
from the CLEO experiment~\cite{CLEOLast}.

\par
We are grateful for the excellent luminosity and machine conditions
provided by our \pep2\ colleagues, 
and for the substantial dedicated effort from
the computing organizations that support \babar.
The collaborating institutions wish to thank 
SLAC for its support and kind hospitality. 
This work is supported by
DOE
and NSF (USA),
NSERC (Canada),
IHEP (China),
CEA and
CNRS-IN2P3
(France),
BMBF and DFG
(Germany),
INFN (Italy),
FOM (The Netherlands),
NFR (Norway),
MIST (Russia), and
PPARC (United Kingdom). 
Individuals have received support from the 
A.~P.~Sloan Foundation, 
Research Corporation,
and Alexander von Humboldt Foundation.


   



\begin{thebibliography}{99}
\bibitem{footnote1}
Charge conjugate decay modes are implicitly included, $\ell =e,\mu$.
\end{thebibliography}
\end{document}